\documentclass[preprint, a4paper, 12pt, oneside, onecolumn]{elsarticle}

\usepackage{amssymb}
\usepackage{amsmath}
\usepackage{booktabs}
\usepackage{microtype}
\usepackage{graphicx}
\usepackage{listings} %
\usepackage{bm}
\usepackage{xcolor}
\usepackage{fancyvrb}
\usepackage{natbib}
\usepackage[T1]{fontenc}

\usepackage{hyperref}
\hypersetup{
    colorlinks=true,
    linkcolor=blue,
    filecolor=magenta,      
    urlcolor=cyan,
}

\newcommand{\fct}[1]{\code{#1()}}
\newcommand{\pkg}[1]{{\fontseries{b}\selectfont #1}}
\newcommand\code{\bgroup\@makeother\_\@makeother\~\@makeother\$\@codex}
\def\@codex#1{{\normalfont\ttfamily\hyphenchar\font=-1 #1}\egroup}
\let\code=\texttt
\let\proglang=\textsf

\let\pkg=\strong
\newcommand{\CRANpkg}[1]{\href{https://CRAN.R-project.org/package=#1}{\pkg{#1}}}

\journal{}
\date{}
\nopreprintlinetrue

\begin{document}

\begin{frontmatter}

\title{Unidimensional and Multidimensional Methods for Recurrence Quantification Analysis with \pkg{crqa}}

 \author[1,2]{Moreno I. Coco}
 \ead{moreno.cocoi@gmail.com}
 \author[3]{Dan M{\o}nster}
 \author[4]{Giuseppe Leonardi}
 \author[5]{Rick Dale}
 \author[6]{Sebastian Wallot}

\cortext[1]{Corresponding author}

 \address[1]{School of Psychology, University of East London, Water Lane, E15 4LZ, London, UK}
  \address[2]{CICPSI, Faculdade de Psicologia, Universidade de Lisboa, Alameda da Universidade, 1649-013 Lisboa, Portugal}
  \address[3]{School of Business and Social Sciences, Aarhus University, DK-8000, Aarhus C, Denmark}
  \address[4]{Institute of Psychology, University of Economics and Human Sciences in Warsaw, 01-043, Warsaw, Poland}
 \address[5]{Department of Communication, University of California, Los Angeles, CA 90005, Los Angeles, USA}
 \address[6]{Department of Language and Literature, Max Planck Institute for Empirical Aesthetics, GR-60322, Frankfurt a.M., Germany}

\begin{abstract}
    \small
      Recurrence quantification analysis is a widely used  method for characterizing patterns in time series. This article presents a comprehensive survey for conducting a wide range of recurrence-based analyses to quantify the dynamical structure of single and multivariate time series, and to capture coupling properties underlying leader-follower relationships. The basics of recurrence quantification analysis (RQA) and all its variants are formally introduced step-by-step from the simplest auto-recurrence to the most advanced multivariate case. Importantly, we show how such RQA methods can be deployed under a single computational framework in R using a substantially renewed version our \pkg{crqa} 2.0 package. This package includes implementations of several recent advances in recurrence-based analysis, among them applications to multivariate data, and improved entropy calculations for categorical data. We show concrete applications of our package to example data, together with a detailed description of its functions and some guidelines on their usage.
\end{abstract}


\end{frontmatter}

\section{Introduction}

In the current article, we present the updated 2.0 version of the \proglang{R} package \pkg{crqa} to perform many variants of recurrence-based analyses \citep{coco2014cross} including some very recent developments for the treatment of multivariate and categorical data. Recurrence-based techniques allow the quantification of temporal structure and generalized autocorrelation properties of individual time series \citep{webber1994dynamical, zbilut1992embeddings}, the quantification of bivariate relationships and coupling between two time series \citep{zbilut1998detecting, marwan2002nonlinear}, as well as the quantification of multidimensional dynamics of multivariate time series \citep{wallot2016multidimensional, wallot2018calculation}. Recurrence-based techniques originate from the description and analysis of dynamical systems \citep{marwan2007recurrence, marwan2002nonlinear}, and have been widely applied to data from physics \citep{alex2015order, ambrozkiewicz2019dynamical, donner2007scale, hilarov2017detection, zolotova2006phase}, physiology \citep{Marwan2002heart, langbein2004visual, thomasson2001recurrence, mestivier2001effects, monster2016physiological, timothy2017classification}, and psychology \citep{abney2014using, coco2018performance, coco2016multilevel, shockley2003mutual,wallot2019switching, wijnants2012interaction}, to name a few fields.

The real success of recurrence-based analyses has revolved around their power of capturing the dynamics of complex and non-stationary time series data, and of time series exhibiting qualitatively different patterns along their temporal evolution \citep{marwan2007recurrence}. This is because recurrence-based analyses are model-free techniques that make few assumptions and hence are well suited for the analysis of complex systems. Moreover, recurrence-based analyses are versatile and can be applied to interval-scale data as well as nominal data, continuously sampled data and inter-event data alike \citep{dale2011nominal,zbilut1998detecting}.

The new version of the \pkg{crqa} package features the integration of major developments in recurrence analysis, such as its extension to multidimensional data, as well as a key simplification of its design and a marked improvement of the underlying computational procedures. It includes useful new functions, including a tool for mining the parameter settings for continuous data, and piece-wise computation of recurrence plots to mitigate computational cost of long time series.

The remainder of this article is divided into two broad sections. In the first section, we provide a concise introduction to the core concepts of recurrence analysis from the simplest case of auto-recurrence of a unidimensional time series, to the most complex case of multidimensional cross-recurrence, which is now integrated in the new version of the \pkg{crqa} package. In the second section, we showcase example applications of the different analysis methods using empirical data, hence providing a hands-on tutorial for how to use the different functions of the package.

\section{Methodological background}

In section \ref{sec:rqa}, we briefly introduce the framework of recurrence quantification analysis (RQA hereafter) with the simplest case of a unidimensional time series. Then, in section \ref{sec:crqa}, we discuss how RQA can be applied to two different unidimensional times series. Finally, in section \ref{sec:mdrqa} and \ref{sec:mdcrqa} we explain how RQA methods can be extended to multidimensional data. 

\subsection{Recurrence quantification analysis (RQA)}
\label{sec:rqa}

The concept of recurrence is at the heart of all recurrence-based analyses \citep{marwan2002nonlinear, trulla1996recurrence}, which mostly apply to time series or sequenced data (but see \cite{wallot2018deriving}). As we will see, recurrences are often defined in terms of phase space coordinates, not directly in terms of the values of a single time series, but the concept is easily demonstrated using this case---a single time series or sequence---as a starting point. Loosely speaking, a recurrence is the repetition of a value in a sequence of data points. More precisely, recurrence in a time series $x$, with $n$ data points $x_1,x_2,\ldots,x_n$ is defined as:

\begin{equation}
    R_{ij} = \left\{
  \begin{array}{lr}
    1 & : x_i = x_j\\
    0 & : x_i \neq x_j
  \end{array}
\right.
\quad
i,j = 1,2, \ldots n
\end{equation}

The above equation defines the elements $R_{ij}$ of the recurrence matrix $\mathbf{R}$ in terms of identical repetitions. Such a definition is useful for a nominal sequence, where there are categorical elements that are either identical or not, and so no meaningful `distance' norm among categories can be defined (see section \ref{sec:rqa:data} for an application to text data). In order to define recurrences for continuous data, we need to establish a threshold parameter (or radius parameter) $\varepsilon$, which provides the width of a tolerance band in the chosen distance norm within which similar, but not identical values in a time series are counted as recurrent:

\begin{equation}\label{eq:Rij_1d}
    R_{ij} = \left\{
  \begin{array}{lr}
    1 & : \vert x_i - x_j \vert \leq \varepsilon \\
    0 & : \vert x_i - x_j \vert > \varepsilon
  \end{array}
\right.
\quad
i,j = 1,2, \ldots n
\end{equation}

Setting a threshold is necessary in most cases for empirical data, because such data feature intrinsic fluctuations as well as measurement error \citep{marwan2007recurrence}.
By using recurrences as defined above, we can convert any unidimensional time series $x$ into a recurrence plot (RP), which is a two-dimensional portrait of its dynamics expressed through its recurrence characteristics.

\begin{figure}
\includegraphics[width=\columnwidth]{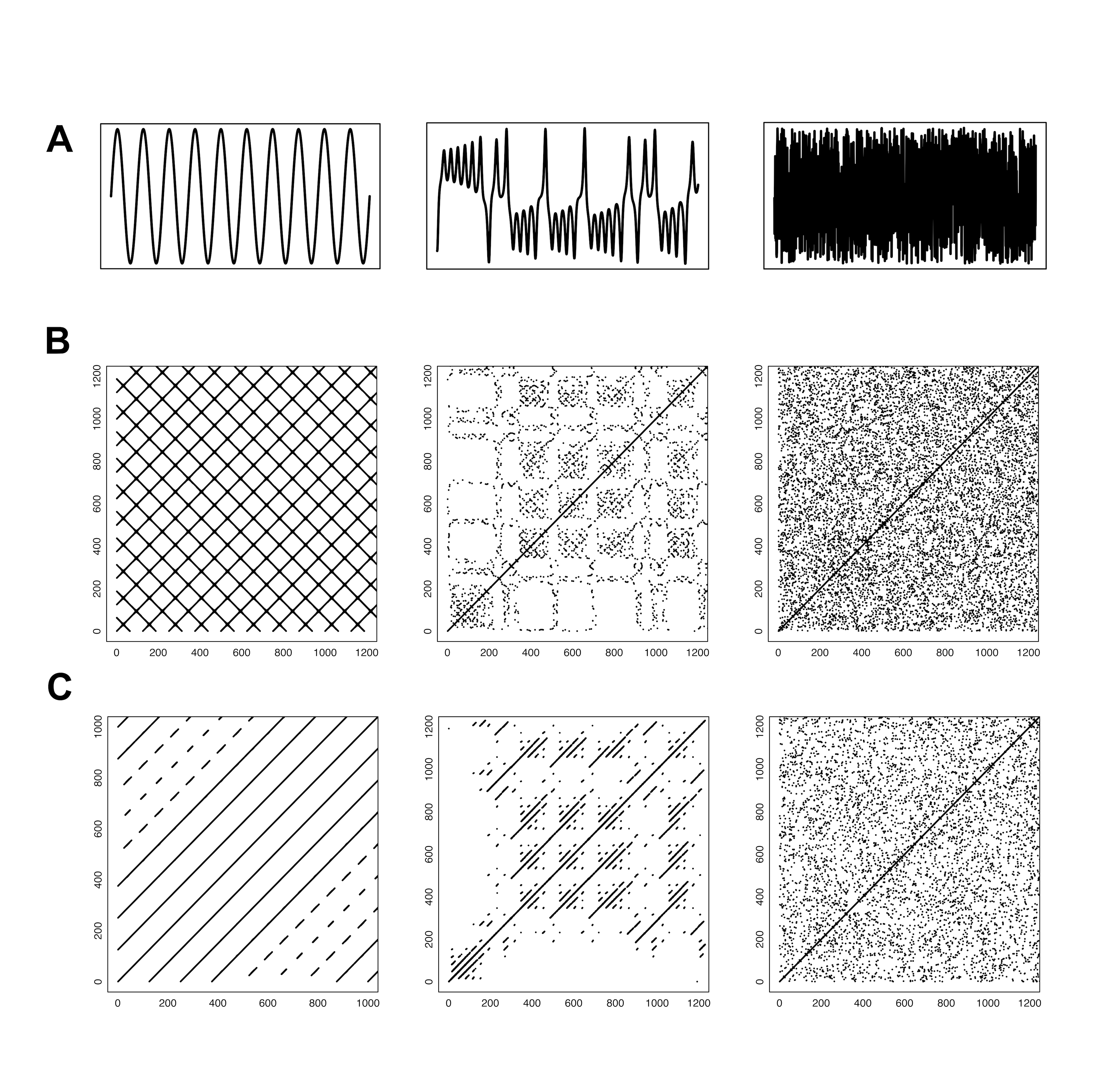}
\caption{\label{fig:examples}The three time series in the first row (A) --- a periodical sine wave, one of the dimensions of the chaotic Lorenz attractor and a white noise signal --- were subjected to recurrence analysis without dimensional embedding, that is the recurrence plots depict recurrences based on the values of the one-dimensional time series (B). The third row (C) shows their associated recurrence plots with dimensional embedding (the sine wave was embedded in 2 dimensions, the Lorenz attractor and the white noise signal in 3 dimensions).}
\end{figure}

Figure~\ref{fig:examples} shows some examples of time series of various complexity (i.e., a sinusoidal, a chaotic attractor and white noise) and their associated RPs, given some value for the threshold parameter $\varepsilon$.

\begin{figure}
\includegraphics[width=\columnwidth]{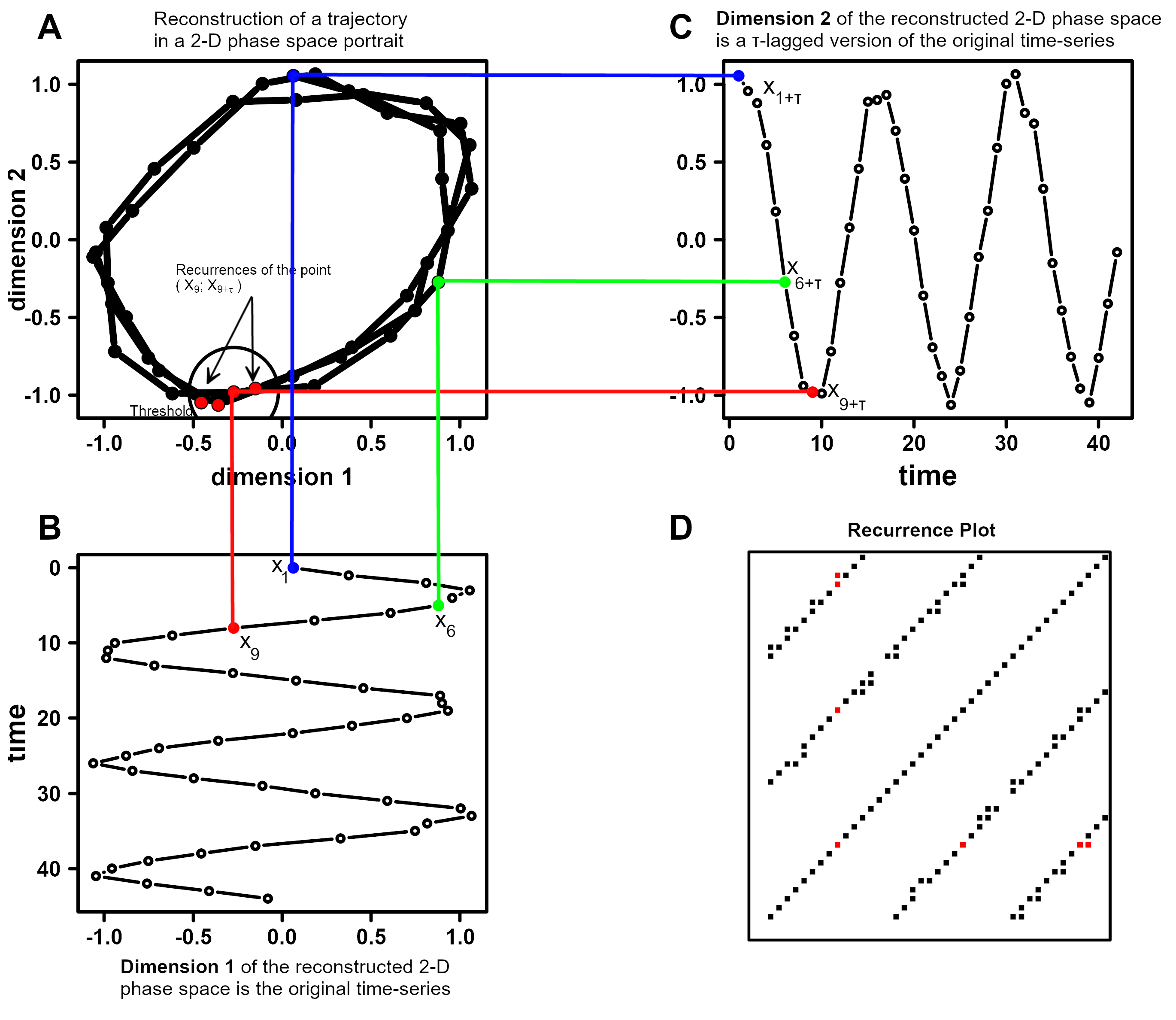}
\caption{\label{fig:embedding}Graphical exemplification of the embedding method. A unidimensional time series (B)--- here represented with time running along the y-axis---is embedded in a two-dimensional phase space (A) by choosing as the second dimension (C) of every point in panel B the $\tau$-lagged value (here $\tau$ = 3) of the very same original unidimensional time series (B). After selecting an appropriate value of the threshold (or radius) parameter (here $\varepsilon$ = 0.25) recurrence analysis can be applied to the trajectory in the two-dimensional phase space (A), and a recurrence plot is generated (D). In the recurrence plot (D), the recurrence points around the coordinates of ($X_9$;$X_{9+\tau}$) are highlighted in red color.}
\end{figure}

If a time series $x$ constitutes the one-dimensional measurement of an underlying multidimensional system, and the dynamics of the underlying dimensions are co-dependent, then these underlying dimensions can be recovered via the method of time-delayed embedding from the unidimensional time series \citep{packard1980geometry, takens1981detecting}. In these cases, the time series $x$ is delayed (or lagged) by a certain number of data points, $\tau$, and the number of times such delays are applied to $x$ depend on an embedding dimension parameter $m$. The time-shifted copies of $x$, $x_{\tau}$, $x_{2\tau}$,\ldots ,$x_{(m-1)\tau}$, can now be integrated into a single $m$-dimensional phase space, which shows the recovered multidimensional dynamics behind the measured unidimensional time series (Figure~\ref{fig:embedding}).

If the time series data comes from a multidimensional system, embedding the data into a higher dimensional phase space before the computation of an RP will improve the quantification of the dynamics of the systems from which that time series was recorded \citep{marwan2007recurrence}. Now, with embedded data, recurrences are defined not in terms of the individual values of the original, unidimensional time series $x$, but in terms of coordinates in $m$-dimensional phase space. It is possible to construct a total of $N = n-(m-1)\tau$ points in phase space, with the following coordinates:

\begin{align}
    \mathbf{X}_1 &= (x_1, x_{1+\tau}, x_{1+2\tau}, \ldots, x_{1+(m-1)\tau}) \nonumber\\
    \mathbf{X}_2 &= (x_2, x_{2+\tau}, x_{2+2\tau}, \ldots, x_{2+(m-1)\tau}) \nonumber\\
    \vdots& \\
    \mathbf{X}_k &= (x_k, x_{k+\tau}, x_{k+2\tau}, \ldots, x_{k+(m-1)\tau}) \nonumber\\
    \vdots&  \nonumber\\
    \mathbf{X}_N &= (x_N, x_{N+\tau}, x_{N+2\tau}, \ldots, x_{n}) \nonumber
\end{align}

In terms of points in the $m$-dimensional phase space, the elements of the recurrence matrix are then given by

\begin{equation}\label{eq:Rij}
    R_{ij} = \left\{
  \begin{array}{lr}
    1 & : \vert\vert \mathbf{X}_i - \mathbf{X}_j \vert\vert \leq \varepsilon \\
    0 & : \vert\vert \mathbf{X}_i - \mathbf{X}_j \vert\vert > \varepsilon
  \end{array}
\right.
\quad
i,j = 1,2, \ldots N
\end{equation}

where $\vert\vert\cdot\vert\vert$ is a distance norm in the $m$-dimensional phase space and $\varepsilon$ is the threshold used to determine whether the points are close enough in phase space to be considered recurring or not. Examples of embedding and the resulting RPs are shown in Figure~\ref{fig:examples}, lower panel, and in Figure~\ref{fig:embedding}, panel D.
Note, however, that in the process of phase space reconstruction, the number of coordinates available in phase space is less than the number of data points in the original time series, the difference being equal to $n-N = \tau (m-1)$. Moreover, the resulting phase space portraits do not exactly reflect the true underlying multidimensional dynamics, but are isomorphic to them \citep{garland2016exploring}.
For continuous time series (i.e., not categorical), the delay $\tau$, the embedding dimension $m$ and the radius $\varepsilon$ are usually unknown, and have to be estimated from the data. $\tau$ can be obtained by examining the average mutual information function (AMI) of $x$ \citep{fraser1986independent}, and the false-nearest-neighbor function (FNN) of $x$, given some value for $\tau$, can in turn be used to determine $m$ \citep{kennel1992determining}. The radius $\varepsilon$  is then chosen to achieve a desired proportion of recurrence points that is expected given the type of data at hand (\cite{webber2005recurrence}; see \cite{wallot2018calculation} for practical information on parameter estimation). The \pkg{crqa} package implements functions to carry out parameter estimation semi-automatically.

RPs are a very useful visualisation to qualitatively explore the dynamics of a time series. However, their main advantage is that they provide the basis to quantify the  dynamics of a time series based on the patterns of recurrence points found in the plot  \citep{zbilut1992embeddings}. There are various measures that can be computed from RPs. Here, we briefly describe a selection of measures, including the most common ones, that are implemented in the new version of the \pkg{crqa} package.

The most basic measure is the recurrence rate (RR) or percentage recurrence, which is defined as the sum of all recurrence points in an RP divided by the area of that RP. RR provides a measure for how many individual values of a time series---or its phase space coordinates---recur over time:

\begin{equation}
    \mathrm{RR} = \frac{1}{N^2} \sum_{i,j=1}^N R_{ij}
\end{equation}

All other measures characterise dynamics by exploiting the patterns of recurrences along the vertical and diagonal structures of the RP (see Table~\ref{tab:rqa_summary} for a concise summary of the measures). In particular, measures based on diagonal-line structures reflect repetitions of the trajectories of the time series, whereas measures based on vertical-line structures focus on the states during which a time series slows down its dynamics. The entropy of the time series can also be computed on the basis of diagonal and vertical line structures of the RP. In version 2.0 of the \pkg{crqa} package, we include a novel entropy measure, which is based on the distribution of the areas of the rectangular structures in an RP generated from categorical time series. This measure provides a more accurate estimation of entropy over the classic diagonal-line entropy for categorical time series that predominantly evolve in terms of changes of states \citep{leonardi2018method}.

\begin{table}[ht]
\renewcommand{\arraystretch}{1.9}
    \centering
    \resizebox{\textwidth}{!}{
    \begin{tabular}{{@{}lll@{}}}\toprule
         \textbf{Measure} & \textbf{Abbreviation} & \textbf{Definition}  \\\midrule
         Recurrence Rate &  RR & $\displaystyle \frac{1}{N^2} \sum_{i,j=1}^N R_{ij}$ \\
         Determinism &  DET & $\displaystyle \left. \sum_{l=l_\mathrm{min}}^N lP(l) \middle/ \sum_{l=1}^N lP(l) \right.$ \\
         Average Diagonal Line Length & L & $\displaystyle \left. \sum_{l=l_\mathrm{min}}^N lP(l) \middle/ \sum_{l=l_\mathrm{min}}^N P(l) \right.$ \\
         Maximum Diagonal Line Length & maxL & $\displaystyle \max(\{l_i\}_{i=1}^{N_l}), \quad N_l = \sum_{l\geq l_\mathrm{min}} P(l)$\\
         Diagonal Line Entropy & ENTR & $\displaystyle  - \sum_{l=l_\mathrm{min}}^N p(l)\log p(l)$ \\
         Laminarity & LAM & $\displaystyle \left. \sum_{v=v_\mathrm{min}}^N vP(v) \middle/ \sum_{v=1}^N vP(v) \right.$\\
         Trapping Time & TT & $\displaystyle \left. \sum_{v=v_\mathrm{min}}^N vP(v) \middle/ \sum_{v=v_\mathrm{min}}^N P(v) \right.$\\
         Categorical Area-based Entropy & catH & $\displaystyle  - \sum_{a>1}^{N_a} p(a)\log p(a)$ \\ 
         \bottomrule
    \end{tabular}%
    }
    \caption{Summary and definition of RQA measures. {Here, $l$ is some diagonal line length on the recurrence plot, i.e. the number of diagonally adjacent recurrence points; $P(l)$ is the histogram or frequency distribution of such diagonal line lengths; $p(l)$ is probability of some diagonal line length; $v$ is some vertical line length on the recurrence plot, i.e., the number of vertical adjacent recurrence points; $P(v)$ is the histogram or frequency distribution of such diagonal line lengths; $l_\mathrm{min}$ and $v_\mathrm{min}$ are the minimum diagonal and vertical line lengths included in the measures ($\ge 2$); $a$ is the value of the area of a rectangular recurrence block generated in a categorical recurrence analysis; $p(a)$ is the probability of the recurrence blocks of area $a$.}}
    \label{tab:rqa_summary}
\end{table}

\subsection{Cross-recurrence quantification analysis (CRQA)}
\label{sec:crqa}

Until now, we focused on quantifying the dynamics of a system by way of recurrences of a single time series. But, the concept of recurrence can be extended to that of cross-recurrence, which extends the univariate recurrence analysis to a bivariate analysis technique that allows quantification of the temporal coupling properties or similarity of two time series \citep{zbilut1998detecting, marwan2007recurrence}. In other words, cross-recurrence is the recurrence of a value in a time series $x$, with data points $x_1, x_2, \ldots ,x_n$ with the values of a time series $y$, with data points $y_1, y_2, \ldots ,y_n$. Formally:

\begin{equation}
     \mathrm{CR}_{ij} = \left\{
  \begin{array}{lr}
    1 & : \vert x_i - y_j \vert \leq \varepsilon \\
    0 & : \vert x_i - y_j \vert > \varepsilon
  \end{array}
\right.
\quad
i,j = 1,2, \ldots n
\end{equation}

As for Equation~\ref{eq:Rij_1d}, a threshold parameter $\varepsilon$ is applied to identify similar, but not necessarily identical, values that are recurrent across the two time series. As in the univariate case, this parameter can be set to values closer to 0, forcing cross-recurrences to be identical values, such as between two nominal (categorical) sequences. Also in the case of cross-recurrence analysis, the two time series $x$ and $y$ can be embedded before computing the cross-recurrence plot (CRP):

\begin{equation}
    \mathrm{CR}_{ij} = \left\{
  \begin{array}{lr}
    1 & : \vert\vert \mathbf{X}_i - \mathbf{Y}_j \vert\vert \leq \varepsilon \\
    0 & : \vert\vert \mathbf{X}_i - \mathbf{Y}_j \vert\vert > \varepsilon
  \end{array}
\right.
\quad
i,j = 1,2, \ldots N
\end{equation}

Commonly it is expected that $x$ and $y$ have the same number of data points, and that the delay and embedding dimension parameters, $\tau$ and $m$, have to be the same too (see \cite{wallot2018calculation,wallot2018analyzing} for practical aspects of parameter estimation)\footnote{It is possible for $x$ and $y$ to be different lengths, and so to produce a rectangular CRP, but this is very rare in practice and it introduces some complications in computing synchrony measures}. The recurrence measures obtained from cross-recurrence quantification analysis are calculated in the same way as for the univariate recurrence quantification analysis (see Table~\ref{tab:rqa_summary}). However, the values for cross-recurrence now reflect the coupled dynamics of the two time series \citep{Shockley2002coupled}, rather than the dynamics of an individual time series in univariate RQA.
There is a key difference between RPs and CRPs. RPs always have recurrence points all along the main diagonal line of the plot (so-called line of identity, LOI), because a time series is by definition recurrent with itself at lag 0, which is what recurrences along the main diagonal reflect (i.e., $x_i=x_j$ when $i=j$). This is not necessarily the case for CRPs as two time series are not necessarily synchronized ($x_i$ need not be the same as $y_j$ when $i=j$). If the time series are not synchronized, cross-recurrences around the main diagonal are absent or sparse. The presence of a full LOI in a CRP implies that the dynamics of the two time series are identical, or that they have a strict linear relationship to each other.

\subsection{Multidimensional recurrence quantification analysis (MdRQA)}
\label{sec:mdrqa}

Multidimensional recurrence quantification analysis (MdRQA) is one of the recent extensions of RQA  \citep{wallot2016multidimensional} that is now also available in the \pkg{crqa} package. MdRQA allows the analysis of multidimensional time series $\mathbf{z}$ with samples $\mathbf{z}_1, \mathbf{z}_2, \ldots , \mathbf{z}_n$, where each point (sample) in $\mathbf{z}$ has $d$ dimensions:

\begin{equation}
    \mathbf{z}_k = (z_{k,1}, z_{k,2}, \ldots , z_{k,d}) 
\end{equation}

Here, recurrences are defined on the $d$-dimensional coordinate space made up of the points of $\mathbf{z}$:

\begin{equation}
    R_{ij} = \left\{
  \begin{array}{lr}
    1 & : \vert\vert \mathbf{z}_i - \mathbf{z}_j \vert\vert \leq \varepsilon \\
    0 & : \vert\vert \mathbf{z}_i - \mathbf{z}_j \vert\vert > \varepsilon
  \end{array}
\right.
\quad
i,j = 1,2, \ldots n
\end{equation}

Like their unidimensional counterparts, multidimensional time series can be embedded into a higher dimensional space. The logic of estimating the delay and embedding dimension parameters, $\tau$ and $m$, are the same as with univariate RQA \citep{wallot2017recurrence,wallot2018analyzing}. However, if one has multivariate time series and wants to estimate embedding parameters for those time series, one should use the multivariate embedding functions which are also provided with the new version of the \pkg{crqa} package, because they provide superior estimates of embedding parameters for multidimensional time series \citep{wallot2018calculation}.

Depending on the underlying data, MdRQA can, in principle, be used for two different purposes. On the one hand, MdRQA can be used to quantify a multidimensional construct, such as physiological arousal, by simultaneously looking at its different measurable dimensions (e.g., heart rate, breathing and body temperature). This would be the multivariate version of how univariate RQA quantifies the dynamics of a single unidimensional time series (e.g., breathing alone). On the other hand, MdRQA can be used to examine the shared dynamics of multiple individual time series, such as, for example, three electrodermal signals measured from three members of a team performing a collaborative task. Here, MdRQA variables would be interpreted as capturing higher-order inter-correlative properties between the three signals at the level of the group \citep{wallot2016multidimensional}.

In either case, one has to make a decision about whether to normalize the different dimensions of the time series or not. If each dimension of the multidimensional time series is normalized, for example $z$-scored, it would effectively give each time series equal weight for the definition of recurrence. In particular, if one does not know how the different time series interact, or if they are measured on different scales without regard to their potential effects on one another, then normalization is recommended, otherwise the risk is to assign a greater weight to the time series bearing greater variance. If one is certain that the values of each dimension of the multidimensional time series are already properly scaled with regard to each other, such as when simultaneously analyzing the three dimensions of the Lorenz system \citep{Lorenz1963b}, then one needs not---and perhaps should not---normalize the dimensions.

\subsection{Multidimensional cross-recurrence quantification analysis (MdCRQA)}
\label{sec:mdcrqa}

Multidimensional cross-recurrence quantification analysis (MdCRQA) extends MdRQA in the same way that CRQA extends RQA. Effectively, MdCRQA allows for the computation of cross-recurrences between two multidimensional time series $\mathbf{x}$ and $\mathbf{y}$ \citep{wallot2019multidimensional}, where cross-recurrences are defined between the two $d$-dimensional coordinate spaces between the points of $\mathbf{x}$ and $\mathbf{y}$:

\begin{align}
    \mathbf{x}_i &= (z_{i,1}, z_{i,2}, \ldots , z_{i,d}) \\
    \mathbf{y}_j &= (z_{j,1}, z_{j,2}, \ldots , z_{j,d}) 
\end{align}

\begin{equation}
    \mathrm{CR}_{ij} = \left\{
  \begin{array}{lr}
    1 & : \vert\vert \mathbf{x}_i - \mathbf{y}_j \vert\vert \leq \varepsilon \\
    0 & : \vert\vert \mathbf{x}_i - \mathbf{y}_j \vert\vert > \varepsilon
  \end{array}
\right.
\quad
i,j = 1,2, \ldots n
\end{equation}

It is important that the different dimensions of the two multivariate time series enter the analysis in the same order. For the example of physiological arousal, if heart rate is the first dimension in $\mathbf{x}$, breathing is the second dimension in $\mathbf{x}$, and body temperature is the third dimension in $\mathbf{x}$, then heart rate also needs to be the first dimension in $\mathbf{y}$, breathing needs to be the second dimension in $\mathbf{y}$, and accordingly body temperature needs to be the third dimension in $\mathbf{y}$. Otherwise, the resulting MdCRQA measure will not be interpretable.

Unidimensional and  multidimensional cross-recurrence analysis can also be performed in a time-dependent manner. This so-called windowed cross-recurrence analysis is conceptually very similar to windowed cross-correlation analysis as in \citet{boker2002windowed} and it can be used to track how cross-recurrence changes over the time course. To that end, one simply partitions the time series of interest into a number of overlapping or non-overlapping sub-series and calculates CRPs for each of the sub-series. For each CRP, i.e., each sub-series of the original time series, the recurrence measures are calculated which allows tracking changes in cross-recurrence over time.

\subsection{The diagonal cross-recurrence profile (DCRP)}
\label{sec:drp}

Finally, from cross-recurrence plots of either two unidimensional time series (i.e., CRQA) or two multidimensional time series (i.e., MdCRQA), it is possible to extract the diagonal cross-recurrence profiles (DCRPs), and use it to capture leader-follower-relationships \citep{dale2011nominal,marwan2007recurrence}. To that end, one has to specify a window size $w$ for the number of lags on the recurrence plot that one wants to investigate. For example, a window-size of 10 data-points, i.e., $w = 10$ would span recurrences of $\pm$ 10 diagonals from the LOI. Specifically, this procedure determines the cross-recurrence rates of each diagonal, $\mathrm{CR}_w$, by summing up all cross-recurrence points that fall along such diagonal, and divide them by its length: 

\begin{equation}
    \mathrm{CR}_{w} = \frac{1}{N - w} \sum_{j-i=w}^{N-w} R_{i,j} 
\end{equation}

\begin{figure}
\includegraphics[width=\columnwidth]{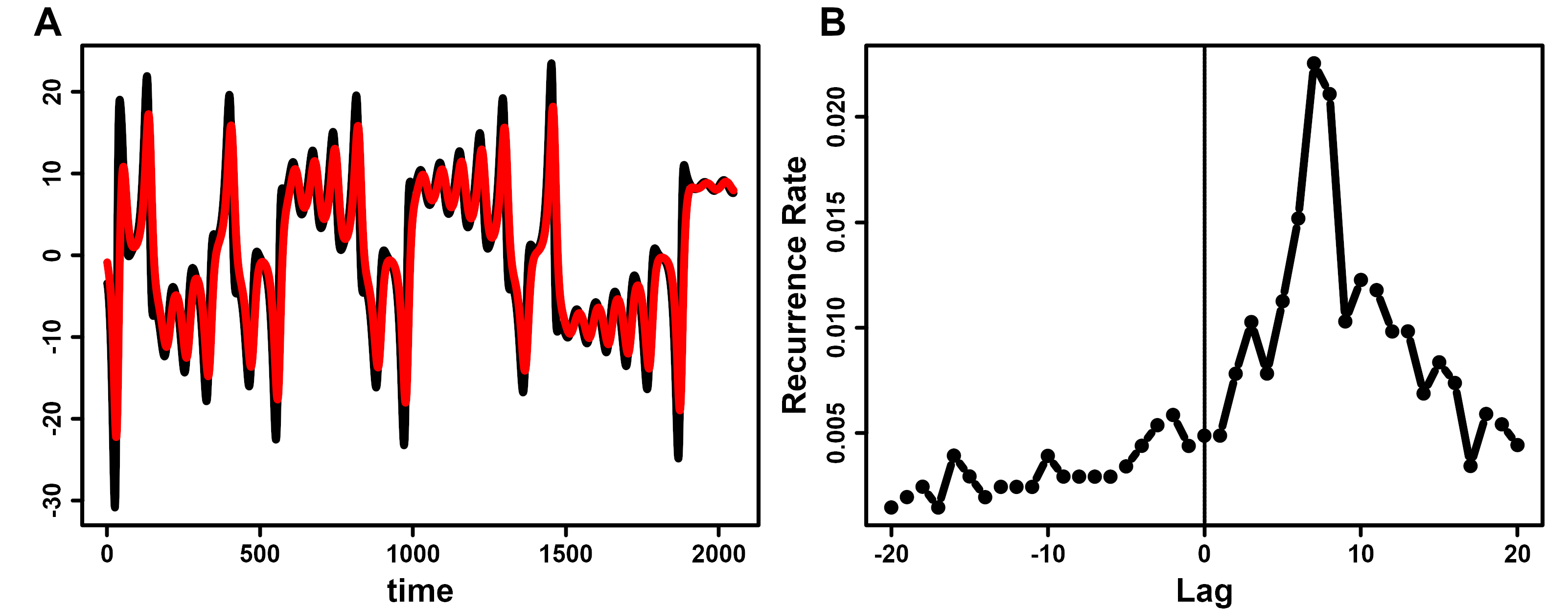}
\caption{\label{fig:dcrp}Illustration of DCRP. Time series of two of the three dimensions from the Lorenz system, one in red, the other one in black (A). After computation of their CRP, one can define their diagonal cross-recurrence profile (DCRP) in order to quantify their time-lagged coupling properties (B). As can be seen, the two time series are most strongly coupled at around the lags of 5 to 9, where a peak in the DCRP can be observed.}
\end{figure}

The DCRP permits quantification of the recurrence rate over different relative lags between two time series. If the peak of cross-recurrences fall along the central diagonal, the line of synchronization (LOS), then this suggests strong coupling at lag~0 between the two time series. If the peak of cross-recurrences falls instead on one of the diagonals off the LOS, it indicates that the dynamics of one time series follow the dynamics of the other time series by some lag equal to that diagonal position (Figure~\ref{fig:dcrp}). 

Note, however that interpreting the lags in terms of the sampling rate of the underlying measured time series only applies to CRPs based on unembedded (often categorical) time series. If the time series are embedded, this means that the observed lag is based on coordinates that are made up of several data points from the respective original time series, and hence introduces a degree of uncertainty with regard to the precise time interval of the lag when one tries to map particular recurrence points back to data points of the original time series. In addition, the leader-follower interpretation of the lags cannot be granted the status of a causal interpretation. For example, a parent can deliberately `lag' their behavior behind that of their child. In such a case, it would be odd to say definitively that the child's behavior is causing the parent's behavior. For this reason, the DCRP has to be interpreted with caution and is best treated as a general description of relative temporal relationships. 

\section{Package design}

The \pkg{crqa} is available from the Comprehensive \proglang{R} Archive Network (\proglang{CRAN}) at \href{https://cran.r-project.org/web/packages/crqa/index.html}{https://cran.r-project.org/web/packages/crqa/index.html}. The software is written in \proglang{R} with the exception of the \code{spdiags} function, which contains a section written in \proglang{Fortran}.
The main function is \code{crqa}, which takes its name from the package. This function performs all types of recurrence quantification analyses discussed in the previous section, and it returns the actual recurrence plot, along with the measures listed in Table~\ref{tab:rqa_summary} extracted from it. Here is how to call this primary function, with arguments and defaults:

\begin{lstlisting}[numbers=none]
crqa(ts1, ts2, delay, embed, rescale = 0, radius, 
    normalize = 0, mindiagline = 2, minvertline = 2, 
    tw = 0, whiteline = FALSE, recpt = NULL, 
    side = 'upper', method = 'crqa', metric = 'euclidean', 
    datatype = 'continuous')
\end{lstlisting}

\begin{table}[!t]
\centering
\resizebox{\textwidth}{!}{
\small
\begin{tabular}{l|p{6.8cm}|p{9cm}}
\hline
Argument & Description &  Usage Notes  \\ \hline
\code{ts1} and \code{ts2} &  The input time series, either unidimensional or multidimensional time series & If auto-recurrence is used (see \code{method = rqa}, then t1 and t2 should be the same time series. \\ 
\code{delay} & A constant indicating the number of time points used to lag the time series & It corresponds to the $\tau$ of the equations.\\
\code{embed} & A constant indicating the number of embedding dimensions applied to the time series  & It corresponds to the $m$ of the equations.\\
\code{rescale} & Whether the distance matrix on which recurrence is evaluated should be rescaled and how & If \code{rescale = 0}, keep the distance matrix as is; if \code{rescale = 1}, rescale the distance matrix to its mean; if \code{rescale = 2}, rescale to distance matrix to its maximum value.\\
\code{radius} & A constant used to decide whether the distance between two points is small enough to be considered recurrent & For categorical time series, the radius needs to be set at values smaller than 1. For continuous time series, the value of the radius needs to tailored to the type of data and its range.\\
\code{normalize} & Normalize the time series & if \code{normalize = 0}, keep the time series at their original scale; if \code{normalize = 1}, normalize the time series to unit interval; if \code{normalize = 2}, z-score the time series. \\
\code{mindiagline} & The minimum number of contiguous points along the diagonals to consider the system into a recurrent state & The default value is usually 2, as it takes a minimum of two points to define any line.\\
\code{minvertline} & The minimum number of contiguous points along the vertical lines to consider the system into a recurrent state & The default value is usually 2, as it takes a minimum of two points to define any line.\\	
\code{tw} & The Theiler window parameter & It defines the number of diagonals off the line of identity that are excluded from recurrence quantification. \\
\code{whiteline} &	 A logical flag to calculate (TRUE) or not (FALSE) empty vertical lines. & The default is FALSE as the calculation of such lines adds on the time of computation. \\
\code{recpt} & A logical flag indicating whether measures of cross-recurrence are calculated directly from a recurrent plot (TRUE) or not (FALSE) & It is mostly useful if the user wants to compute joint-recurrence analysis. The RP or CRP is supplied in place of ts1, and ts2 needs to be assigned as NA.\\ 
\code{side} & A string indicating the side of the recurrence plot on which recurrence measures should be calculated & For \code{side = upper}, recurrence measures are calculated on the upper triangle of the RP, for \code{side = lower} on the lower triangle of the RP, for \code{size = both} on the full RP. Note, the line of identity is automatically excluded for \code{upper} and \code{lower} setting.\\
\code{method} &	A string to indicate the type of recurrence analysis to perform & For \code{method = rqa}, Auto-recurrence is calculated, i.e., a unidimensional series; for \code{method = crqa} cross-recurrence is calculated; for \code{method = mdcrqa} multidimensional recurrence is calculated. Note, the default value is \code{crqa}.\\
\code{metric} &	A string to indicate the type of distance metric used & To see the list of all other possible metrics see the help for the \fct{rdist} function. Note, the default is \code{euclidean}.\\
\code{datatype} & A string (continuous or categorical) to indicate the nature of the data type & If the time series contain categorical information, it will automatically be recoded into a continuous integer-based time series with a warning sent to the user.\\
\hline
\end{tabular}
}
\caption{\label{tab:crqa:arguments} Overview of the arguments that can be used in \code{crqa} to set up the recurrence quantification analysis.}
\end{table}

Several arguments in this function allow the user to refine aspects of the computation. For example, the user can modify the metric to obtain the distance matrix when estimating recurrence, or the settings of thresholds to accept contiguous points as recurring (see Table~\ref{tab:crqa:arguments} for the list of arguments of \fct{crqa} with a brief explanation of each). \fct{drpfromts} can be used to obtain the diagonal cross-recurrence profile, and this function is built using the \fct{crqa} at its core. In fact, most arguments stay exactly the same, and we will illustrate the additional arguments that are specific to this function when describing it. Likewise, the functions \fct{wincrqa} and \fct{windowdrp} are built on the main function \fct{crqa} and are used to compute windowed cross-recurrence. The package also contains functions, such as \fct{optimizeParam}, to automatically estimate the settings for the three main parameters of RQA analyses, i.e., radius $\varepsilon$, delay $\tau$ and embedding dimension $m$, for continuous measures. As recurrence quantification analysis is heavy on memory requirements, the package features a function (\fct{piecewiseRQA}) which can be used to break down the analysis of long time series into smaller and more manageable chunks that are computationally more tractable. Finally, the package provides the user with functions to simulate data from classic examples of dynamical systems, such as the Lorenz (\fct{lorenzattractor}) or categorical series with different distributions (\fct{simts}). In what follows, we briefly describe the data available with the package. These data are used to illustrate the different functionalities of the \fct{crqa} package.

\subsection{Related Packages}

A search of CRAN shows very few other packages offering alternative methods to compute recurrence quantification analysis. In particular, \CRANpkg{tseriesChaos}, \CRANpkg{nonlinearTseries} and \CRANpkg{RHRV}. The function \code{recurr} in \CRANpkg{tseriesChaos} computes a recurrence plot for a (section of) a single unidimensional timeseries, but it does not compute any derived measures from the plot characterising the dynamics of the system, nor does it handle cross recurrence plots or any of the many extensions to simple recurrence plots provided by our \CRANpkg{crqa} package. Going a little further, the function \code{rqa} in \CRANpkg{nonlinearTseries} returns key measures from the recurrence plots (e.g., recurrence rate), as well as, a quick visualisation of the recurrence plot (available also with the function \code{RecurrencePlot}). Finally, the function \code{RecurrencePlot} in \CRANpkg{RHRV} is simply a wrapper built on the function in \CRANpkg{nonlinearTseries} with the same name, but specifically tailored to  electrocardiogram data. The functions available in \CRANpkg{nonlinearTseries} provide very basic functionality in terms of the range of metrics available and optimization routines. They do not allow the user to examine diagonal structures for leader-follower analyses, nor explore the evolution of recurrence rate using windowed methods.  All such features are integrated into \pkg{crqa}, which, to the best of our knowledge, is the most comprehensive statistical package to perform recurrence quantification analysis in R. 

\subsection{Data}
\label{sec:data}

Different types of categorical and continuous time series, both unidimensional and multidimensional, are available with the package. The command \code{load(crqa)} will load the data into the \proglang{R} workspace. In particular, we include a nursery rhyme ``The wheels on the bus'' by Verna Hills to illustrate the most basic recurrence quantification analysis. This text is a vector of 120 strings (i.e., the words of the song), and as it is extremely simple and highly repetitive, it makes it a very good example to illustrate the core concept of recurrence. Then, we move on to cross-recurrence quantification analysis and include in the data object of the package eye-tracking data from the study by \cite{richardson2005looking}. In this study, a narrator describes the characters of a TV series (Friends) to a listener, who will have to later answer some comprehension questions about them, while their eye-movement are co-registered. Here, we use a single trial of this study, which is stored as a dataframe of 2,000 observations of six possible screen locations that are looked at by the narrator and the listener. These are numerically coded from 1 to 6, representing a 2x3 visual grid\footnote{There are two more states, 10 and 11, to indicate when the listener or the narrator blinked or looked outside of the screen, or to identify possible blinks. They are coded with a different number so that these two states will not recur when radius is set near 0.}). This data will also be used to illustrate diagonal and windowed cross-recurrence. Finally, we illustrate multidimensional cross-recurrence analysis using hand movement data from the study by \cite{wallot2016beyond}. In this study, dyads were instructed to cooperate in a complex LEGO joint construction task, under different conditions, while their hand movements and heart-rates were co-registered. Again, we select only a single trial of hand-movement from the turn-taking condition. The dataframe comprises of 5,799 observations from two participants (P1 and P2) for the dominant (\_d) and non-dominant (\_n) hand. 

\subsection[Using the crqa package]{Using the \pkg{crqa} package}

\subsubsection{RQA}
\label{sec:rqa:data}

As explained in section \ref{sec:rqa}, RQA entails computing the auto-recurrence of a unidimensional time series. In the context of the nursery rhyme, we expect clear phases of recurrence to emerge because the words greatly repeat. 
In order to run this analysis, we use the main function \code{crqa}, and specify in the argument \code{method} that we are running an RQA analysis (\code{method = "rqa"}). As the data that we use is categorical, we need to specify in the argument \code{datatype} that the nature of the data is categorical (\code{datatype = "categorical"}) this will automatically recode the categorical states of the series (i.e., the words) into unique numerical integers so that recurrence can be computed using a \code{radius} that has to be smaller than 1 (e.g., \code{radius = 0.01}), so that we can capture the recurrence of identical words. For categorical RQA, the delay and embedding dimension have to be set to 1 (\code{delay = 1; embed = 1\footnote{If embedding dimension is set to higher values, this becomes equivalent to doing recurrence on $n$-grams, where $m=n$. In fact this interpretation of categorical recurrence creates bridges to traditional natural language processing, summarized in \cite{dale2018dynamic}}.}). We also need to set the Theiler window parameter to 1 (\code{tw} = 1), so that we can exclude the LOI from all recurrence measures. Finally, the same unidimensional time series has to be input both as \code{ts1} and \code{ts2} to obtain its auto-recurrence.  

\begin{lstlisting}[numbers = none]
res <- crqa(text, text, delay = 1, embed = 1, rescale,
            radius = 0.01, normalize, mindiagline, minvertline,
            tw = 1, whiteline, recpt, side, method = "rqa",
            metric = "euclidean", datatype = "categorical")
\end{lstlisting}

We can use the plotting function \code{plotRP} to visualise the resulting recurrence plot. This function provides some basic arguments to change the size of the points in the plot (\code{pcex}), the color (\code{cols}), or their type (\code{pch}) which are taken verbatim from the generic \code{plot} function.

\begin{lstlisting}[numbers=none]
RP <- res$RP
parC <- list(unit = 10, labelx = "Time", labely = "Time",
cols = "black", pcex = .5, pch = 15, las = 0,
labax = seq(0, nrow(RP), 10),
labay = seq(0, nrow(RP), 10)) 
plotRP(RP, parC)
\end{lstlisting}

\begin{figure}
\includegraphics[width=\columnwidth]{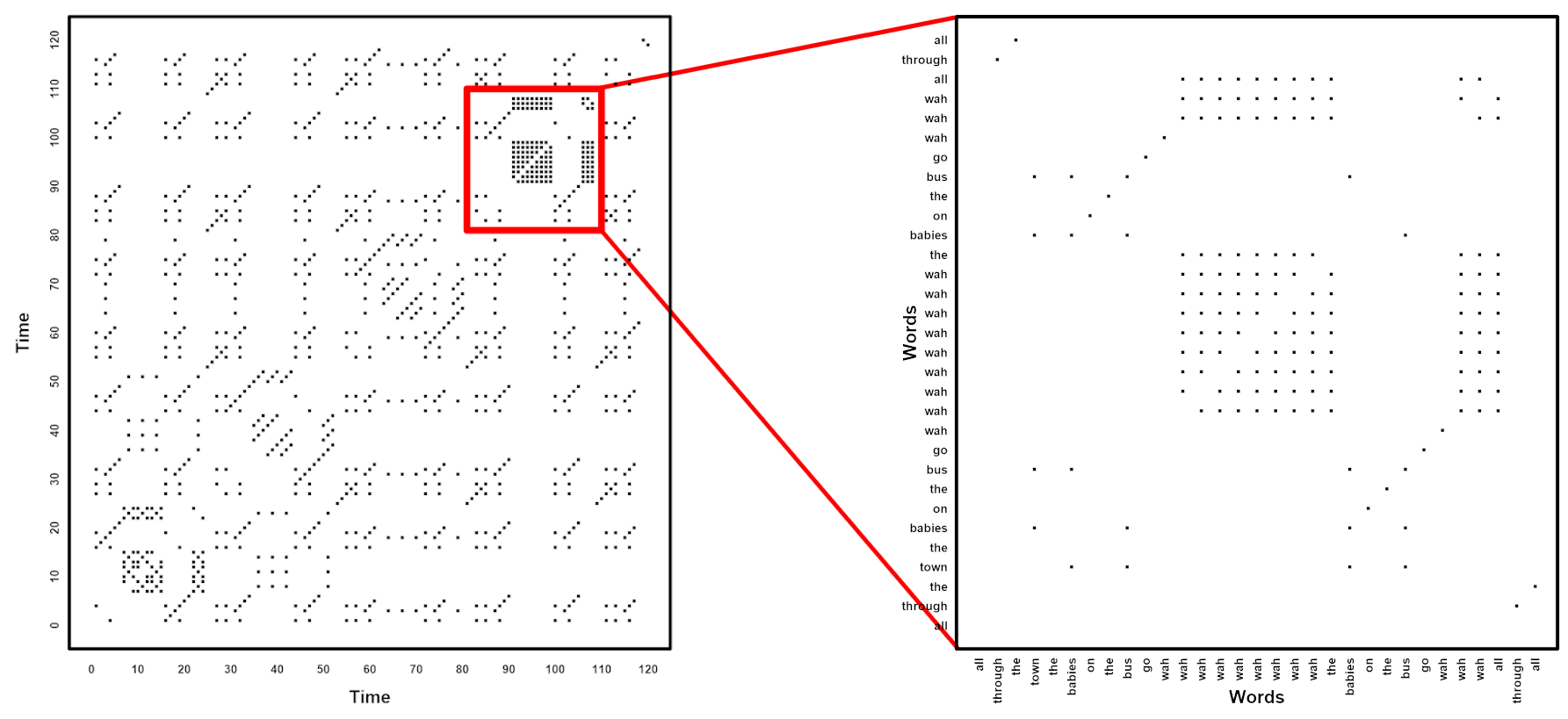}
\caption{\label{fig_rqa_text} Recurrence quantification of the nursery rhyme `wheels on the bus'. On the left panel, we show the full recurrence and the red box indicates the region of RP that is zoomed in on the right panel.}
\end{figure}

In order to get a closer understanding of recurrences, we zoom into a segment of the text, re-run the \code{crqa()} function and visualise it (Figure~\ref{fig_rqa_text}). We add the labels of the axes (x,y), print the words vertically using the \code{las} argument, and decrease the temporal \code{unit} argument to print each individual word on the axes.

\begin{lstlisting}[numbers=none]
text_zoom <- text[81:110]
ans_zoom <- crqa(text_zoom, text_zoom, delay, embed,
                rescale, radius, normalize, mindiagline,
                minvertline,tw, whiteline, recpt, side,
                method, metric, datatype)

RP <- ans_zoom$RP
parC$labay <- parC$labax <- text_zoom
parC$las <- 2; parC$unit <- 1
parC$labelx <- parC$labely <- "Words"
plotRP(RP, parC)
\end{lstlisting}

 As it can be clearly seen in Figure~\ref{fig_rqa_text}, the bulk of recurrence in that portion is driven by the repeated use of the single word \textit{wah}. When looking at a few measures associated with the RP, we observe an overall determinism of 85.4\%, which implies that the system is fairly repetitive, an average diagonal line length of 3.88, which implies that on average there are sequences of four words that repeat, and a maximum diagonal length of 9, which means that the longest sequence repeating is made of 9 words. Some measures such as determinism or the average diagonal line length depend on the setting of the argument \code{mindiagline} (equivalent to $l_\text{min}$ in Table~\ref{tab:rqa_summary}). The default value of this argument is 2, as two contiguous points form a line, but it can depend on the type of data (e.g., words vs.\ eye-movement), or the sample rate at which it is acquired (55~Hz vs.\ 1,000~Hz). For example, if we have acquired data at 1,000~Hz, we would practically have one data-point every 2~ms. This means that if we use a default \code{mindiagline} of 2, we would be considering as lines, any states that contiguously repeat over a 4~ms window. This value would certainly be unrealistic for some type of responses that unfolds over longer period of time (e.g., an eye-movement fixation lasts for an average of 200~ms). 

\subsubsection[CRQA: crqa()]{CRQA: \code{crqa()}}

\begin{figure}
\includegraphics[width=\columnwidth]{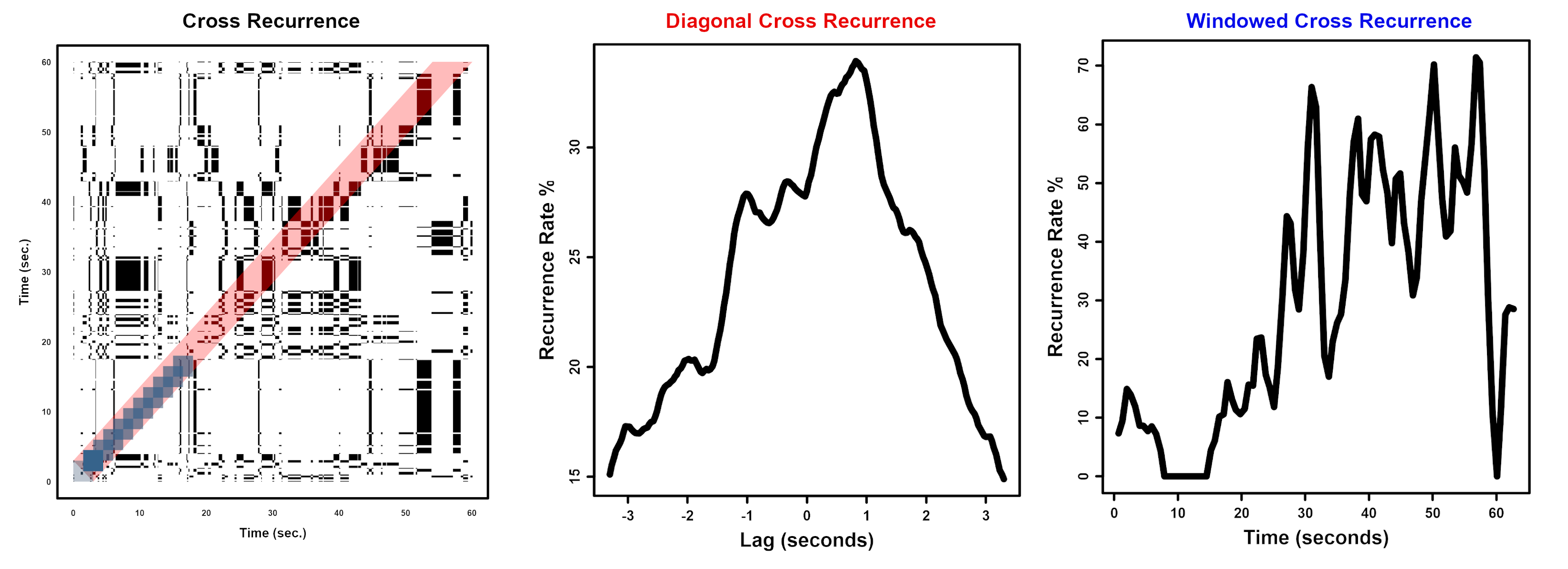}
\caption{\label{fig:crqaeye} Cross-recurrence using a single trial of eye-movement data measured on a listener and a narrator. On the left panel, we visualise the full cross-recurrence plot. The transparent red band represents the lags around the line of coincidence that have been used to calculate the diagonal cross-recurrence plot (center panel), whereas the blue transparent squares within it are the overlapping windows used to compute the windowed cross-recurrence profile (right panel).}
\end{figure}

As already explained in section~\ref{sec:crqa}, cross-recurrence is an extension of auto-recurrence to two different unidimensional time series. In order to run a cross-recurrence analysis with the \pkg{crqa} package, we simply need to change the \code{method} argument to \code{method = "crqa"}. Here, we illustrate its use through two time series of eye-movement data explained in section \ref{sec:data}. Also, we input two different time series of eye-movement data, \code{narrator} and \code{listener}, rather than just one. An optional argument that is available in the \code{crqa} function is the \code{side} (upper, lower or both) of the recurrence plot on which recurrence measures are computed. This may be useful, for example, for researchers interested in leader-follower dynamics (see Figure~\ref{fig:crqaeye}, left panel, for the visualisation of the cross recurrence plot).  

\subsubsection[Diagonal-CRQA: drpfromts]{Diagonal-CRQA: \code{drpfromts}}

In section~\ref{sec:crqa}, we explained what diagonal cross-recurrence is, and how it can be used. In the \code{crqa} package, this measure is computed by the function \code{drpfromts}, which utilises the same arguments of the main \code{crqa} function plus an additional argument, \code{windowsize}, to define the number of lags (or diagonals) around the line of synchronization (LOS) of the CRP over which recurrence rate is computed. In the example visualised in Figure~\ref{fig:crqaeye}, center panel,  we have chosen a window of 100 lags, which spans about $\pm 3$~seconds around the LOS. We can clearly see that the peak recurrence is shifted by $\approx 1$~second from the LOS, i.e., lag 0. This reflects the time taken by the listener to look at the same panel the narrator was looking at---namely, about 1 second for a listener to ``catch up'' to the speaker.

\begin{lstlisting}[numbers=none]
res <- drpfromts(narrator, listener, windowsize = 100, 
                radius = 0.001, delay = 1, embed = 1, 
                rescale = 0, normalize = 0,
                mindiagline = 2, minvertline = 2, tw = 0, 
                whiteline = F, recpt = F, side = "both",  
                method = "crqa", metric = "euclidean", 
                datatype = "continuous")
\end{lstlisting}

\subsubsection[Windowed-CRQA: windowdrp]{Windowed-CRQA: \code{windowdrp}}

Windowed cross-recurrence captures the evolution of recurrence rate over time. In the context of the eye-movement data, this measure reflects how consistently listener and narrator are looking at the same scene location at any given point in time. More importantly, the windowed methodology reveals how recurrence changes over time (refer to section~\ref{sec:drp} for more details). We use the function \code{windowdrp} to compute this measure, which again shares the same arguments of \code{crqa}, plus three more that are specific to it. In particular, we have to set: (a) the size of the window that slides over the time-course, e.g., \code{windowsize =  100}, (b) the step that we want this window to move, e.g., \code{windowstep = 20}, and (c) the number of lags\footnote{Note, that the number of lags cannot be greater than the size of the window.} within the window of interest over which recurrence rate is computed, e.g., \code{lagwidth = 50}. In Figure~\ref{fig:crqaeye}, right panel, we observe that recurrence rate grows over time, which means that the narrator and the listener tend to look more and more at the same panels as the trial progresses. 

\begin{lstlisting}[numbers=none]
res <- windowdrp(narrator, listener, windowstep = 20,
                windowsize = 100, lagwidth = 50, 
                radius = 0.001, delay = 1, embed = 1, 
                rescale = 0, normalize = 0, mindiagline = 2, minvertline = 2, tw = 0,  whiteline = F, 
                side = "both", method = "crqa",  
                metric = "euclidean", datatype = "continuous")
\end{lstlisting}

\subsection{MdCRQA}

Finally, we compute multidimensional cross-recurrence by setting the \code{method} argument, i.e., \code{method = "mdcrqa"}. Note, in order to compute multidimensional recurrence, the user needs to provide the same dataframe as input to both \code{ts1} and \code{ts2}. This method is also available for \code{drpfromts} and \code{windowdrp}. Applied to the hand movement data, we first restructure the data of the two participants into independent sets. We re-use the same parameter settings of \cite{wallot2016beyond} to compute MdCRQA, and leave all other arguments with their default values. 

\begin{lstlisting}[numbers=none]
P1 <- cbind(handset$P1_TT_d, handset$P1_TT_n) 
P2 <- cbind(handset$P2_TT_d, handset$P2_TT_n)

res <- crqa(P1, P2, delay = 5, embed = 2, 
            rescale = 0, radius = 0.1, normalize = 0, 
            mindiagline = 10, minvertline = 10, tw = 0, 
            whiteline, recpt, side, method = "mdcrqa", 
            metric, datatype)
\end{lstlisting}

\subsection[Breaking down the computation: piecewiseRQA]{Breaking down the computation: \code{piecewiseRQA}} 

Often, researchers are interested in time series which contain several thousands of observations. Sometimes the dimensionality of these time series can be reduced without losing too much information, such as by down-sampling. This strategy may not always be possible or serve the researcher's purpose. Recurrence quantification analysis can require more RAM than is available in standard laptops or personal workstations, hence making it nearly impossible to run. In the new version of the \pkg{crqa} package, we provide the user with the \code{piecewiseRQA} function, which can be used to compute all different variants of recurrence quantification analysis described above on long time series. Conceptually, this function divides the time series into blocks, it obtains a recurrence plot for each individual block, and then fills the original recurrence plot in with all such sub-blocks, before computing the measures.\footnote{This function is similar to the \code{crp\_big} function in the long-standing CRP-toolbox for MATLAB by Marwan and colleagues: http://tocsy.pik-potsdam.de.} 

\begin{figure}
\includegraphics[width=\columnwidth]{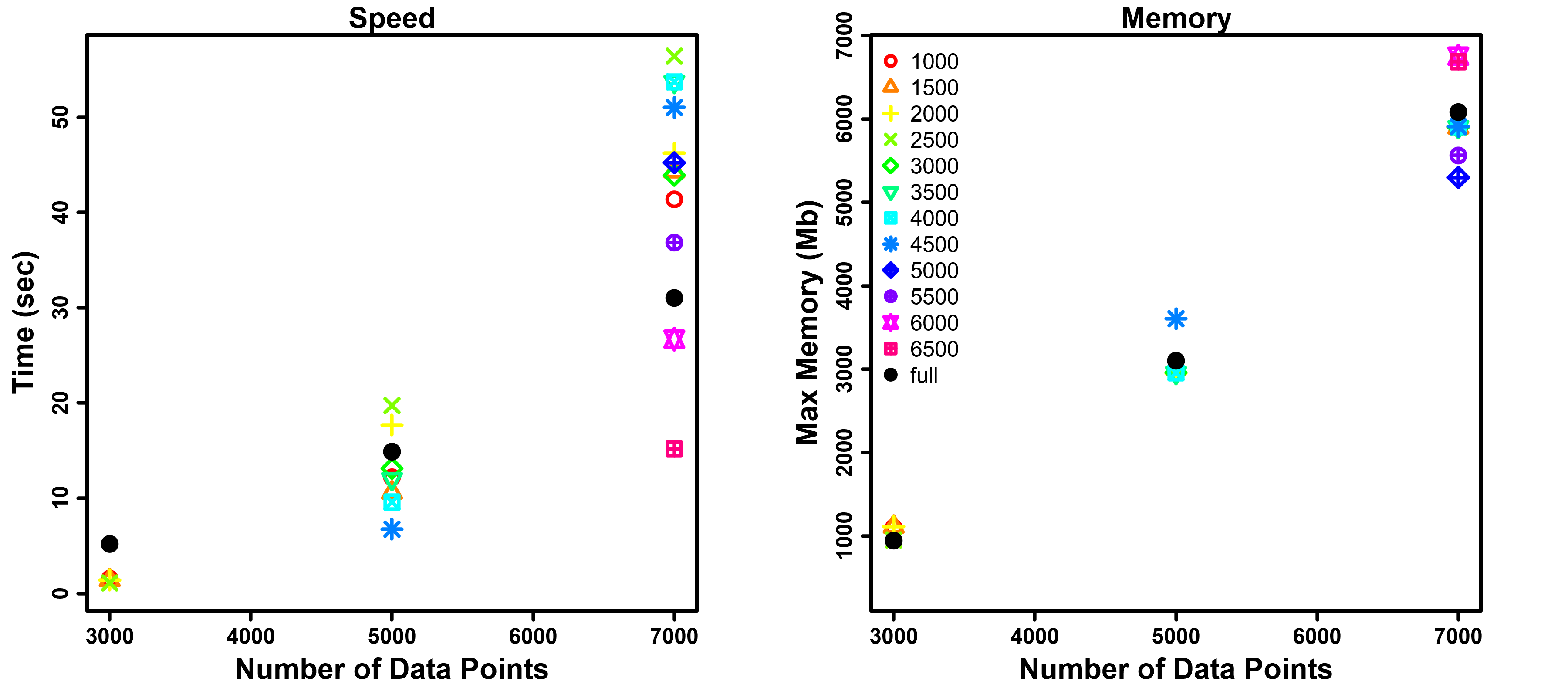}
\caption{\label{fig:performance} Evaluating the speed (time in seconds, left panel) and memory (peak RAM in MB, right panel) performance of crqa() and piecewiseRQA() for simulated data of increasing size (from 3000 to 7000 data points). We compare the use of blocks of different sizes (from 1000 to 6500 in increments of 500, coded using color and point type) with the case of running crqa() on the entire sequence of data points.}
\end{figure}

For example, if we are handling a time series of 10,000 observations, we could divide it into 10 blocks of 1,000 observation each. \code{piecewiseRQA} has exactly the same arguments that we have already encountered in the main \code{crqa} function, but has an additional two that are used to control the size of the block, e.g., \code{blockSize = 100}, and the argument \code{typeRQA}, which can take two options, either \code{full} or \code{diagonal}. If the value for \code{typeRQA} is \code{diagonal} only the diagonal cross-recurrence will be computed; if \code{full}, the recurrence measures will be obtained out of the full plot\footnote{Currently, the windowed cross-recurrence is not implemented to work with the \code{piecewiseRQA}.}. In Figure~\ref{fig:performance}, we visualise the computational speed (left panel) and memory demand (right panel) on simulated time series of sinusoids of increasing size, and compare what happens if we run the \code{piecewiseRQA}, also with blocks of increasing size, as compared to running the main \code{crqa} function. We can clearly see that for time series of increasing length, memory demands are kept lower by the \code{piecewiseRQA} function as compared to the \code{crqa}. However, we can also see that there is a wide variance for blocks of different sizes. Therefore, it may be wise to explore the block sizes to find the one that can optimize the computational performance over a single trial before running the piecewise recurrence analysis on an entire dataset.

\subsection[Estimating starting parameters: optimizeParam]{Estimating starting parameters: \code{optimizeParam}}

The last function we showcase is \code{optimizeParam}, which helps the user exploring the space of values for the parameters of delay, embedding dimension and radius to compute recurrence quantification on continuous valued time series. In particular, \code{optimizeParam} first estimates the average mutual information (AMI) of either the unidimensional or multidimensional time series, and chooses the value that minimizes it. Then, it takes such a delay value, and evaluates the embedding dimensions maximizing the false-nearest neighbors (FNN)\footnote{Users can also access the functions to compute delay (\code{MdDelay}) and embedding dimensions (\code{MdFnn}) independently of \code{optimizeParam}.}. As a last step, \code{optimizeParam} applies the values of delay and embedding dimension obtained, to find a radius which returns a recurrence rate within a minimum and maximum value established by the user. We apply \code{optimizeParam()} to simulated sinusoids to show the unidimensional case, and to the hand movement data of \cite{wallot2016beyond} to show the multidimensional case.

In order to set up the estimation of the delay and embedding dimensions for unidimensional time series, the user has to decide how to choose the value of average mutual information (i.e., \code{typeami = mindip}, the lag at which minimal information is observed, or \code{typeami = maxlag}, the maximum lag at which minimal information is observed) and the relative percentage of information gained in FNN, relative to the first embedding dimension, when higher embeddings are considered (i.e., \code{fnnpercent}). Then, as \code{crqa} is integrated in the \code{optimizeParam} to estimate the radius, most of the arguments are the same (e.g., \code{mindiagline} or \code{tw}), except the number of values of that are considered (i.e., \code{radiusspan = 100}).

\begin{lstlisting}[numbers=none]
ts1 <- seq(0.1, 200, .1)
ts1 <- sin(ts1) + linspace(0, 1,length(ts1))
ts2 <- ts1
par <- list(method = "rqa", metric = "euclidean", 
            maxlag =  20, radiusspan = 100, normalize = 0, 
            rescale = 4,  mindiagline = 10, minvertline = 10, 
            tw = 0, whiteline = FALSE, recpt = FALSE, 
            side = "both", datatype = "continuous", 
            fnnpercent  = 10,  typeami = "mindip")
results <- optimizeParam(ts1, ts2, par, 
                         min.rec = 2, max.rec = 5)
print(unlist(results))
\end{lstlisting}

\begin{lstlisting}[numbers=none]
 radius emddim  delay 
 0.17   2       18 
\end{lstlisting}

For multidimensional series, the user needs to specify the right RQA method (i.e., \code{method = "mdcrqa"}). Then, for the estimation of the delay via AMI: (1) \code{nbins} the number of breaks used to define the bins within which the two-dimensional histogram (or frequency distribution) of the original and delayed time series are computed and (2) the \code{criterion} to select the delay (\code{firstBelow} to use the lowest delay at which the AMI function drops below the value set by the \code{threshold} argument, and \code{localMin} to use the position of the first local AMI minimum). The estimation of the embedding dimensions instead needs the following arguments: (1) \code{maxEmb}, which is the maximum number of embedding dimensions considered, (2) \code{noSamples}, which is the number of randomly drawn coordinates from phase space used to estimate the percentage of false-nearest neighbors, (3) \code{Rtol}, which is the first distance criterion for separating false neighbors, and (4) \code{Atol}, which is the second distance criterion for separating false neighbors. The radius is estimated as before. 

\begin{lstlisting}[numbers=none]
par$method = "mdcrqa"; par$nbins = 50 
par$criterion = "firstBelow"; par$threshold = 1.6
par$maxEmb = 20; par$numSamples = 500; par$Rtol = 10; 
par$Atol = 2

results <- optimizeParam(P1, P2, par, 
                         min.rec = 2, max.rec = 5)
print(unlist(results))
\end{lstlisting}

\begin{lstlisting}[numbers=none]
radius  emddim  delay 
0.032   11      2
\end{lstlisting}

\section{Conclusion}

This paper describes recurrence quantification analysis, a statistical method to characterise the nonlinear dynamics of a system. It has received a growing interest by researchers across very different fields from physiology to psychology because of its flexibility, ease of application, and explanatory power. In particular, we explain recurrence analysis from the simplest case of auto-recurrence of a unidimensional time series to the most complex case of multidimensional cross-recurrence. More importantly, we presented a significantly updated version of the \pkg{crqa} to perform all different variants of recurrence analysis described in the theoretical section of this manuscript. We showcased the different functions available in \pkg{crqa} with real data of categorical and continuous nature, and illustrated how starting parameters for continuous data can be obtained (i.e., radius, embedding dimension and delay) as well as handling long time series in a memory efficient way using additional functions available in \pkg{crqa}.  

It is useful to end on some observations regarding the broader relevance of the package presented here. The RQA methodology and the updated \code{crqa} tap into a number of evolving problems in data analysis across various disciplines. For example, there is a drive to improve measures and models of multimodal data \citep{abawajy2015comprehensive,dale2015integrative,lahat2015multimodal}, including multi-person measures  \citep{cooke2013interactive,lopez2017combining,schilbach2013toward,von2016verbal,wallot2016beyond}. RQA and its multidimensional counterpart implemented in our package constitute an extraordinarily expansive analysis tool for exploring varied kinds of complex and multidimensional data. In addition, a demand for more dynamic quantitative analyses has now also penetrated into the social sciences \citep{chemero2011radical,friedenberg2009dynamical,spivey2008continuity,ward2002dynamical}. \code{rqa} is designed to be a comprehensive analysis package for studying the dynamics of diverse systems, especially systems that exhibit high degrees of interdependence, and that show signatures in their dynamics that are critical for understanding them. 

\subsection{Acknowledgments}

SW acknowledges support from the Deutsche Forschungsgemeinschaft (DFG, German Research Foundation)--WA 3538/4-1. MIC acknowledges support from the Funda\c{c}\^ao para a Ci\^encia e Tecnologia under grant agreement PTDC/PSI-ESP/30958/2017 

\newpage

\bibliography{crqa-references}
\bibliographystyle{elsarticle-num-names}

\newpage

\end{document}